\magnification=1200
%\magnification=1000
%\magnification=\magstep1
%\vsize=20truecm
%\voffset=1.00truein
\settabs 18 \columns
%\hoffset=3.75truecm
%\hoffset=1.00truein
%\hsize=14truecm

%\nopagenumbers
%\baselineskip=17 pt
\baselineskip=12 pt
%\ifnum\pageno=1
\topinsert \vskip 1.25 in
\endinsert
%\vsize=7.5in
%\fi

%\def\mybox{\sqcap\kern-.66em\sqcup\kern.65em}
\def\sqr#1#2{{\vcenter{\vbox{\hrule height.#2pt
 \hbox{\vrule width.#2pt height#1pt \kern#1pt
 \vrule width.#2pt} \hrule height.#2pt}}}}

\def\operp{\hbox{${\kern+.25em{\bigcirc}
\kern-.85em\bot\kern+.85em\kern-.25em}$}}
%\def\gapprox
%{\hbox{$
%\smash{lower0.5ex\hbox{$\scriptstyle>$}} \atop
%\smash{raise0.3ex\hbox{$\scriptstyle \sim$}}
%$}}
%\def\lapprox
%{\hbox{$
%\smash{lower0.5ex\hbox{$\scriptstyle<$}} \atop

%\smash{raise0.3ex\hbox{$\scriptstyle \sim$}}
%$}}
\def\lsim{\;\raise0.3ex\hbox{$<$\kern-0.75em\raise-1.1ex\hbox{$\sim$}}\;}
\def\gsim{\;\raise0.3ex\hbox{$>$\kern-0.75em\raise-1.1ex\hbox{$\sim$}}\;}
\def\no{\noindent}

\def\ce{\centerline}
\def\ve{\vfill\eject}
\def\rdots{\mathinner{\mkern1mu\raise1pt\vbox{\kern7pt\hbox{.}}\mkern2mu
 \raise4pt\hbox{.}\mkern2mu\raise7pt\hbox{.}\mkern1mu}}

\def\e e{$e^+ e^-$ }

%\input epsf

%End of Beginning Formats
%Beginning of Letter Heading

\rightline{UCLA/00/TEP/14}
\rightline{January 2000}
\vskip1.0cm

\ce{{\bf $SL_q(3)$ Fields}}
\vskip.5cm
\ce{\it R. J. Finkelstein}
\vskip.3cm
\ce{Department of Physics and Astronomy}
\ce{University of California, Los Angeles, CA  90095-1547}
\vskip1.0cm

\no {\bf Abstract.}  The $q$-field theories are constructed by
substituting quantum groups for the usual Lie groups.  In earlier papers
this construction was carried out for the quantum group $SU_q(2)$.  Here
the investigation is extended to $SL_q(3)$.  The resulting theory describes
two sectors, one sector lying close to the standard theory and accessible
by perturbation theory, while the second sector describes particles that
should be difficult to detect and become invisible in the $q=1$ limit.
In this note we discuss these hypothetical particles: three quark-like
spinor particles coupled to three gluon-like vector particles.
\vskip 1.0cm

\no PACS numbers 81R50, 81T13.
\ve

\no {\bf 1.}~~ Since the Lie groups may be considered as degenerate forms
of the quantum groups, it may be of interest to generalize the symmetry of a
conventional field theory by replacing its Lie group by the corresponding
quantum group.  When this generalization is carried out, it is found that
the state space must be expanded to describe additional degrees of freedom
that are absent in the corresponding Lie theory.

Our earlier work was based on the simple $SL_q(2)$ quantum group.$^1$  Since the
phenomenology of elementary particles requires Lie groups of higher
rank, we shall here extend the earlier discussion of $SL_q(2)$ to describe
fields that lie in the $SL_q(3)$ algebra.
\vskip.5cm

\no {\bf 2.}~~In general the quantum groups may be defined by the relations$^2$
$$
RT_1T_2 = T_2T_1R \eqno(2.1)
$$
\no where
$$
\eqalignno{T_1 &= T\otimes I & (2.2) \cr
T_2 &= I\otimes T & (2.3) \cr}
$$
\no and the $R$ matrix satisfies the Yang-Baxter equation
$$
R_{12}R_{13}R_{23} = R_{23}R_{13}R_{12}~. \eqno(2.4)
$$
\no Written out, (2.1) reads
$$
\sum^n_{m,p=1} R_{ij,mp} t_{mk}t_{p\ell} =
\sum^n_{m,p=1} t_{jp}t_{im}R_{mp,k\ell} \eqno(2.5)
$$
\no where for the series $A_{n-1}$
$$
R = q\sum^n_{i=1} e_{ii}\otimes e_{ii} + \sum^n_{\scriptstyle i,j=1\atop
\scriptstyle i\not= j}
e_{ii}\otimes e_{jj} + (q-q_1) \sum^n_{\scriptstyle i,j=1\atop i>j}
e_{ij}\otimes e_{ji}~. \eqno(2.6)
$$
\no Here
$$
[e_{mn}]_{ij} = \delta_{mi}\delta_{nj} \quad \hbox{and} \quad
q_1=q^{-1}~. \eqno(2.7)
$$

For this choice of $R$ Eq. (2.5) becomes
$$
\eqalign{qt_{ik}t_{j\ell}\delta_{ij} &+ t_{ik}t_{j\ell}(1-\delta_{ij}) +
(q-q_1)t_{jk}t_{i\ell}\theta(i-j) \cr
&~~=qt_{jk}t_{i\ell}\delta_{k\ell} + t_{j\ell}t_{ik}(1-\delta_{k\ell})
+(q-q_1)t_{jk}t_{i\ell}\theta(\ell-k) \cr}
$$
\no where
$$
\eqalign{\theta(s) &=1 \quad s>0 \cr
&= 0 \quad s\leq 0~. \cr}
$$
\no By (2.8) one finds the following special cases, depending on whether
the two matrix elements are in the same row $(i=j)$:
$$
\ell < k \quad t_{i\ell}t_{ik} = qt_{ik}t_{i\ell} \eqno(2.9)
$$
\no the same column $(k=\ell)$:
$$
j < i \quad t_{ik}t_{jk} = qt_{jk}t_{ik} \eqno(2.10)
$$
\no neither same row nor same column
$$
\eqalignno{i<j,k<\ell \quad t_{ik}t_{j\ell} &= t_{j\ell}t_{ik} +
(q-q_1)t_{jk}t_{i\ell} & (2.11) \cr
i<j,k>\ell \quad t_{ik}t_{j\ell} &= t_{j\ell}t_{ik} & (2.12) \cr}
$$
\no These results may be summarized as follows:  Choose any rectangle in
the full $N\times N$ matrix, say
$$
\left(\matrix{ik & i\ell \cr jk & j\ell \cr}\right) \quad \hbox{in} \quad
\left(\matrix{t_{11}~\ldots~t_{1N} \cr t_{N1}~\ldots~t_{NN} \cr}\right)
\eqno(2.13)
$$
\no Then ($i<j$ and $k<\ell$)
$$
\eqalignno{&t_{ik}t_{i\ell} = qt_{i\ell}t_{ik} & (2.14) \cr
&t_{ik}t_{jk} = qt_{jk}t_{ik} & (2.15) \cr
&(t_{ik},t_{j\ell}) = (q-q_1)t_{jk}t_{i\ell} & (2.16) \cr
&(t_{i\ell},t_{jk}) = 0 & (2.17) \cr}
$$
\no i.e. the 4 vertices exhibited in (2.13) belong to 
the algebra of $GL_q(2)$.  Note that
all the commuting elements lie on lines of positive slope.  Therefore the
maximum set of commuting elements lie on the minor diagonal.

There is also the quantum determinant
$$
\hbox{det}_qT = \sum_\sigma (-q)^{\ell(\sigma)}
t_{1\sigma_1}\ldots t_{n\sigma_n} \qquad
\sigma \epsilon ~\hbox{symm}(n) \eqno(2.18)
$$
\no where $\ell(\sigma)$ is the number of inversions in the permutation 
$\sigma$.

The quantum determinant commutes with all elements of $T$
$$
(\hbox{det}_qT,t_{ij}) = 0~. \eqno(2.19)
$$
\vskip.5cm

\line{{{\bf 3.}~~The quantum group $GL_q(3)$.} \hfil}
\vskip.3cm

Let us introduce the following notation
$$
T = \left(\matrix{E_+(1,1) & E_+(1,2) & H(1) \cr
E_+(2,1) & H(2) & E_-(2,3) \cr
H(3) & E_-(3,2) & E_-(3,3) \cr} \right) \eqno(3.1)
$$
\no Consider a matrix realization of the elements of $T$ where the
commuting set $\{H\}$ is Hermitian:
$$
H_i = \bar H_i~. \eqno(3.2)
$$
\no By (2.14) and (2.15)
$$
\eqalignno{E_+H &= qHE_+ & (3.3) \cr
E_-H &= q_1HE_- & (3.4) \cr}
$$
\no where $E$ amd $H$ share either a row or column index.  By Hermitian
conjugation of (3.3) and (3.4)
$$
\eqalignno{H\bar E_+ &= \bar q\bar E_+ H & (3.5) \cr
H\bar E_- &= \bar q_1\bar E_-H & (3.6) \cr}
$$
Therefore we may set
$$
q=\bar q \quad \hbox{and} \quad \bar E_+ = E_-~. \eqno(3.7)
$$
\no Then
$$
T = \left(\matrix{\bar E(1) & \bar E(2) & H(1) \cr
\bar E(3) & H(2) & E(2) \cr H(3) & E(3) & E(1) \cr} \right)~.
\eqno(3.8)
$$
\no Note that $E(1)$ and $\bar E(1)$ may be interchanged if $q$ is replaced
by $1/q$.  We also have
$$
\eqalign{(E(2),E(3)) &= 0 \cr
(H(1),E(3)) &= 0 \cr
(H(3),E(2)) &= 0 \cr
(H(i), H(j)) &= 0 \qquad i,j = 1,2,3  \cr} \eqno(3.9)
$$
\vskip.5cm

\line{{{\bf 4.}~~Representation of the algebra.} \hfil}
\vskip.3cm

Introduce the basis states $|n_i\rangle$, eigenstates of the $H_i$.  Then
$$
\eqalignno{E_+H_i|n_i\rangle &= qH_iE_+|n_i\rangle & (4.1) \cr
\noalign{\hbox{or}}
H_i(E_+|n_i\rangle) &= q_1h_i(n_i)~~(E_+|n_i\rangle)~. & (4.2) \cr}
$$
\no Define $|n_i+1\rangle$ by
$$
E_+|n_i\rangle = \lambda(n_i)|n_i+1\rangle~.
\eqno(4.3)
$$
\no Then
$$
H_i|n_i+1\rangle = q_1h_i(n_i)|n+1\rangle \eqno(4.4)
$$
\no or the eigenvalues of $H_i$ are given by
$$
\eqalignno{h_i(n_i+1) &= q_1h_i(n_i) & (4.5) \cr
\noalign{\hbox{and}}
h_i(n_i) &= q_1^{n_i}h_i(0) & (4.6) \cr}
$$
\no Likewise
$$
\eqalign{H(E_-|n\rangle) &= qh(n)(E_-|n\rangle) \cr
&= h(n-1)(E_-|n\rangle) ~. \cr} \eqno(4.7)
$$
\no Therefore set
$$
\eqalignno{E_-|n_i\rangle &= \mu(n_i)|n_i-1\rangle & (4.8) \cr
E_-|0\rangle~&= 0~. & (4.9) \cr}
$$
\no In the above equations if
$$
\eqalign{E_+~~\hbox{is}~~\bar E(1)~,~~\hbox{then}~~
H~~\hbox{is}~~H(1)~~\hbox{or}~~H(3) \cr
E_+~~\hbox{is}~~\bar E(2)~,~~\hbox{then}~~
H~~\hbox{is}~~H(1)~~\hbox{or}~~H(2) \cr
E_+~~\hbox{is}~~\bar E(3)~,~~\hbox{then}~~
H~~\hbox{is}~~H(2)~~\hbox{or}~~H(3) \cr}
\eqno(4.10)
$$
\no The complete notation for the basis states is
$$
|\vec n\rangle = |n_1,n_2,n_3\rangle \eqno(4.11)
$$
\no where $(n_1,n_2,n_3)$ refer to $(H_1,H_2,H_3)$.  In this notation
(4.8) reads
$$
\eqalignno{E_-(2)|n_1,n_2,n_3\rangle &= \mu_2(n_1,n_2,n_3)
|n_1-1,n_2-1,n_3\rangle & (4.12) \cr
E_-(3)|n_1,n_2,n_3\rangle &= \mu_3(n_1,n_2,n_3)
|n_1,n_2-1,n_3-1\rangle & (4.13) \cr
E_-(1)|n_1,n_2,n_3\rangle &= \mu_1(n_1,n_2,n_3)
|n_1-1,n_2-2,n_3-1\rangle & (4.14) \cr
E_-(2)|0,0,n_3\rangle &= 0 \qquad E_-(1)|0,n_2,0\rangle = 0 & (4.15) \cr
E_-(3)|n_1,0,0~\rangle &= 0 & (4.15) \cr}
$$
\no In this representation $E(1)$ behaves differently from $E(2)$ and
$E(3)$ since
$$
(\bar E(1),H(2)) = \tilde q~\bar E(2)\bar E(3)~, \quad
\tilde q = q-q_1~. \eqno(4.16)
$$
\no There is no corresponding equation for $\bar E(2)$ or $\bar E(3)$.
Then (4.16) implies
$$
\eqalign{
&[h(n_2)-h(n_2^\prime)]\langle\vec n^\prime|\bar E(1)|\vec n\rangle
=\tilde q\langle\vec n^\prime|\bar E(2)\bar E(3)|\vec n\rangle \cr
&= \tilde q\langle n_1+1,n_2+2,n_3+1|\bar E(2)|n_1,n_2+1,n_3+1\rangle
\langle n_1,n_2+1,n_3+1|\bar E(3)|n_1,n_2,n_3\rangle~. \cr} \eqno(4.17)
$$
\no where $\vec n^\prime = (n_1+1,n_2+2,n_3+1)$.
Of the three $E$-operators only $E(1)$ moves all three quantum numbers. 
The raising and lowering operators (determining transition amplitudes) are subject
to simple selection rules that may be found as follows:

For the $E$ and $H$ shown in (3.8)
$$
\eqalignno{&\langle n|\bar EH|m\rangle = q\langle n|H\bar E|m\rangle &
(4.18) \cr
&\langle n|\bar E|m\rangle [h(m)-qh(n)] = 0 & (4.19) \cr
&\langle n|\bar E|m\rangle[q_1^m-q_1^{n-1}] = 0 & (4.20) \cr
&\langle n|\bar E|m\rangle = 0~, \quad m\not= n-1 & (4.21) \cr}
$$
\no Similarly
$$
\langle n|E|m\rangle = 0 \qquad m\not= n+1~. \eqno(4.22)
$$

These selection roles are shown in (4.12) and (4.13) for $E(2)$ and
$E(3)$.  There is an additional selection rule shown in (4.14) and (4.17)
for $E(1)$.
\vskip.5cm

\line{{{\bf 5.}~~Restrictions on the Transition Amplitudes
and the Value of $q$.} \hfil}
\vskip.3cm

The transition amplitudes are restricted by the following relations
$$
\eqalignno{(\bar E(1),~E(1)) &= \tilde q~H(1) H(3) \qquad
\tilde q = q-q_1 & (5.1) \cr
(\bar E(2),~E(2)) &= \tilde q~H(1)H(2) & (5.2) \cr
(\bar E(3),~E(3)) &= \tilde q~H(2)H(3) & (5.3) \cr}
$$

Since these relations have a common structure we first discuss only (5.1).
The diagonal element of this equation may be written as follows:
$$
\langle n_1n_2n_3|(\bar E(1),E(1))|n_1n_2n_3\rangle =
\tilde q\langle n_1n_2n_3|H(1)H(3)|n_1n_2n_3\rangle \eqno(5.4)
$$
\no where $|n_1n_2n_3\rangle$ is a common eigenstate of
$H(1),H(2)$, and $H(3)$.

Eq. (5.4) becomes
$$
\sum_{\vec p}|\langle\vec p|E(1)|\vec n\rangle|^2 -
\sum_{\vec p}|\langle\vec n|E(1)|\vec p\rangle|^2 =
\tilde q~h^{(n_1)}(1)h^{(n_3)}(3)~. \eqno(5.5)
$$
\no If $\vec n=(n_1,n_2,n_3)$, then by (4.14)
$$
\eqalign{\langle\vec p|E(1)|\vec n\rangle &= 0 \quad \hbox{unless} \quad
\vec p = (n_1-1,n_2-2,n_3-1) \cr
\langle\vec n|E(1)|\vec p\rangle &= 0 \quad \hbox{unless} \quad
\vec p = (n_1+1,n_2+2,n_3+1)~. \cr} \eqno(5.6)
$$
\no Hence, denoting the lowest eigenvalue of $H_i$ by $\alpha_i$ we have
$$
\eqalign{|\langle n_1&-1,n_2-2,n_3-1|E(1)|n_1,n_2,n_3\rangle|^2 -
|\langle n_1,n_2,n_3|E(1)|n_1+1,n_2+2,n_3+1\rangle|^2 \cr 
&=\tilde q\alpha_1\alpha_3q_1^{n_1}q_1^{n_3} \cr} \eqno(5.7)
$$
\no or
$$
f(n_1-1,n_2-2,n_3-1)-f(n_1,n_2,n_3)=g(n_1,n_3) \eqno(5.8)
$$
\no where
$$
\eqalignno{f(n_1,n_2,n_3) &= |\langle n_1,n_2,n_3|E(1)|n_1+1,n_2+2,n_3+1\rangle|^2 & (5.9) \cr
g(n_1,n_3) &= \tilde q\alpha_1\alpha_3q^{-n_1-n_3} & (5.10) \cr}
$$

Set $n_1=n_2=n_3=0$ in (5.8).  Then
$$
f(-1,-2,-1)-f(0,0,0) = g(0,0)~.
$$
\no But $f(-1,-2,-1)=0$ since $|000\rangle$ is the vacuum state.  Then
$$
f(0,0,0) = -g(0,0) = -\tilde q\alpha_1\alpha_3~. \eqno(5.11)
$$
\no By (5.8)
$$
f(n_1+1,n_2+2,n_3+1) = f(n_1,n_2,n_3)-g(n_1+1,n_3+1)
$$
\no and
$$
f(n_1+m,n_2+2m,n_3+m) = f(n_1,n_2,n_3) - \sum^m_1 g(n_1+s,n_3+s)~. \eqno(5.12)
$$
  
Set $n_1=n_2=n_3=0$.  Then
$$
\eqalignno{f(m,2m,m) &= f(0,0,0) - \sum^m_1 g(s,s) & (5.13) \cr
&= -\tilde q\alpha_1\alpha_3\langle m\rangle_{q^{-2}} & (5.14) \cr}
$$
\no by (5.10) where
$$
\langle m\rangle_{q^{-2}} = {q^{-2m}-1\over q^{-2}-1}~. \eqno(5.15)
$$
\no By (5.9) and (5.14)
$$
|\langle n,2n,n|E(1)|n+1,2(n+1),n+1\rangle|^2 =
-\tilde q\alpha_1\alpha_3\langle n\rangle_{q^{-2}} \eqno(5.16) 
$$

The corresponding discussion for $E(2)$ and $E(3)$ is simpler since only
two quantum numbers change when $E(2)$ and $E(3)$ act, while all three
$(n_1~n_2~n_3)$ are changed when $E(1)$ acts.  The results for $E(2)$
and $E(3)$ are as follows:
$$\eqalignno{
|\langle n,n,n|E(2)|n+1,n+1,n\rangle|^2 &= -\tilde q\alpha_2\alpha_1
\langle n\rangle_{q^{-2}} & (5.17) \cr
|\langle n,n,n|E(3)|n,n+1,n+1\rangle|^2 &= -\tilde q\alpha_2\alpha_3
\langle n\rangle_{q^{-2}}~. & (5.18) \cr}
$$

Since the left sides of (5.16)-(5.18) are positive one has
$$
-\tilde q\alpha_i\alpha_j>0 \qquad i,j = 1,2,3 ~~~ i\not= j \eqno(5.19)
$$
\no The preceding equations (5.19) imply that $\alpha_i\alpha_j$ always
has the same sign.  Therefore the $\alpha_i$ are also all of the same sign,
the $\alpha_i\alpha_j$ are positive, and
$$
-\tilde q > 0 
$$
\no or by (5.1)
$$
q < 1~. \eqno(5.20)
$$

\ve

\line{{{\bf 6.}~~ The Quantum Determinant.} \hfil}
\vskip.3cm

The magnitudes of the transition amplitudes may be restricted by the
quantum determinant.  The quantum determinant corresponding to (3.8) and
computed according to (2.18) is
$$
\eqalign{\Delta = \bar E(1) H(2) E(1) &-q\bigl[\bar E(2)\bar E(3) E(1) +
\bar E(1) E(2) E(3)\bigr] \cr
&-q^3 H(3)H(2)H(1) \cr
&+q^2\bigl[\bar E(2) E(2) H(3) + H(1)\bar E(3) E(3)\bigr]~. \cr}
\eqno(6.1)
$$
\no Since $\Delta$ belongs to the center of the algebra it may be set
equal to unity:
$$
\Delta = 1~. \eqno(6.2)
$$

Applied to the $H$-vacuum we have
$$
\Delta|0\rangle = |0\rangle \eqno(6.3)
$$
\no implying
$$
-q^3 h_o(3) h_o(2)
h_o(1) = 1~.  \eqno(6.4)
$$
\no Let us again set
$$
h_o(i) = \alpha_i~. \eqno(6.5)
$$
\no Then
$$
\alpha_1\alpha_2\alpha_3 = -q_1^3~. \eqno(6.6)
$$

Since $\alpha_1,\alpha_2$ and $\alpha_3$ all have the same sign by
(5.19), they are all negative.  

In general
$$
\langle\vec n|\Delta|\vec p\rangle = \langle\vec n|\Delta|\vec n\rangle
\delta(\vec n,\vec p)~. \eqno(6.7)
$$
\no To evaluate $\langle\vec n|\Delta|\vec n\rangle$ we need
$$
\langle\vec n|\bar E(1) H(2) E(1)|\vec n\rangle =
\alpha_2 q_1^{n_2-2}|\langle\vec p_1|E(1)|\vec n\rangle|^2 \eqno(6.8)
$$
\no and
$$
\langle\vec n|\bar E(2)\bar E(3) E(1) + \bar E(1)E(2)E(3)|\vec n\rangle =
2\alpha_2 q_1^{n_2-1}|\langle\vec p_1|E(1)|\vec n\rangle|^2~. \eqno(6.9)
$$
\no To obtain (6.9) we have used (6.8) and
$$
\tilde q E(2)E(3) = (H(2),E(1))~. \eqno(6.10)
$$
\no Then
$$
\eqalign{\langle\vec n|\Delta|\vec n\rangle = &-\alpha_2q_1^{n_2-2}
|\langle\vec p_1|E(1)|\vec n\rangle|^2-q^3\alpha_1\alpha_2\alpha_3
q^{n_1+n_2+n_3} \cr
&+q^2\bigl[\alpha_3q_1^{n_3}|\langle\vec p_2|E(2)|\vec n\rangle|^2 +
\alpha_1q_1^{n_1}|\langle \vec p_3|E(3)|\vec n\rangle|^2\bigr] \cr} \eqno(6.11)
$$
\no or by (6.3)
$$
\eqalign{1-q_1^{n_1+n_2+n_3}=q^2\{\alpha_1q_1^{n_1}|\langle \vec p_3|E(3)|\vec n\rangle|^2 &+ \alpha_3q_1^{n_3}|\langle\vec p_2|E(2)|\vec n\rangle|^2 \cr
&-\alpha_2q_1^{n_2}|\vec p_1|E(1)|\vec n\rangle|^2\} \cr} \eqno(6.12)
$$
\no where
$$
\eqalign{\vec p_1 &= (n_1-1,n_2-2,n_3-1) \cr
\vec p_2 &= (n_1-1,n_2-1,n_3) \cr
\vec p_3 &= (n_1,n_2-1,n_3-1) \cr} \eqno(6.13)
$$
\no Note that the curly bracket in (6.12) is negative since $q_1>1$.

If $\vec n = (n,n,n)$ then by (6.12) and (5.16)-(5.18)
$$
1-q_1^{3n} = q^2(-\tilde q\alpha_1\alpha_2\alpha_3)
q_1^n\langle n\rangle_{q_1^2} \eqno(6.14)
$$
\no or
$$
q_1^n=1 \eqno(6.15)
$$
\no by (6.6).

Since $q_1$ is real, $n=0$.  It follows that there are no solutions of
(6.2) for any state $|nnn\rangle$ except the vacuum state $|000\rangle$.
Therefore the restriction on the $q$-determinant acts as an exclusion
principle forbidding any state $|nnn\rangle$ except $|000\rangle$ in which
the quantum numbers of the three $H$-particles agree.
\vskip.5cm

\line{{{\bf 7.}~~Invariant Forms.} \hfil}
\vskip.3cm

Let us restate (2.5) to make explicit the distinction between $T_1$ and
$T_2$
$$
\sum_{mp} R_{ijmp}T(1)_{mk}
T(2)_{p\ell} = \sum_{mp}
T(2)_{jp} T(1)_{im}
R_{mpk\ell}~. \eqno(7.1)
$$
\no Then
$$
\eqalign{\sum_{\scriptstyle mp\atop \scriptstyle k\ell}R_{ijmp}
T(1)_{mk} T(2)_{p\ell}
(&T(2)^{-1})_{\ell s} (T(1)^{-1})_{kt}\cr 
&=\sum_{\scriptstyle mp\atop \scriptstyle k\ell}
T(2)_{jp} T(1)_{im} R_{mpk\ell}
(T(2)^{-1})_{\ell s}
(T(1)^{-1})_{kt} \cr} \eqno(7.2)
$$
\no or
$$
R_{ijts} = \sum_{\scriptstyle mp\atop \scriptstyle k\ell}
T(2)_{jp} T(1)_{im} R_{mpk\ell}
(T(2)^{-1})_{\ell s}(T(1)^{-1})_{kt}~.
\eqno(7.3)
$$
\no Consider two fields $\psi(x)$ and $\chi(x)$ and the following
associated bilinear
$$
\eqalign{\sum\psi_{ij}(x)R_{ijts}\chi_{ts}(x) &=
\sum(\psi_{ij}(x)T(2)_{jp}T(1)_{im})
R_{mpk\ell}(T(2)^{-1})_{\ell s}
(T(1)^{-1})_{kt}\chi_{ts}(x)) \cr
&= \sum\psi^\prime_{mp}(x)R_{mpk\ell}\chi^\prime_{k\ell}(x) \cr} \eqno(7.4)
$$
\no by (7.3), where
$$
\eqalignno{\psi^\prime_{mp} &= \sum \psi_{ij}T(2)_{jp}
T(1)_{im} & (7.5) \cr
\chi^\prime_{k\ell} &= \sum(T(2)^{-1})_{\ell s}
(T(1)^{-1})_{kt} \chi_{ts} & (7.6) \cr}
$$
\no Then $\psi R\chi$ is a bilinear invariant under the $T$-transformations
(7.5) and (7.6).

There is also an invariant trilinear interaction stemming from the
invariance of the quantum determinant.  To show this, rewrite (2.18) as
follows:
$$
\Delta^q = \sum_{j_1\ldots j_n} (-q)^{-\ell(i_1\ldots i_n)}
t_{i_1j_1}\ldots t_{i_nj_n}(-q)^{\ell(j_1\ldots j_k)} \eqno(7.7)
$$
\no where we count the number of inversions between $(i_1\ldots i_n)$ and
$(j_1\ldots j_n)$ rather than the number of inversions between
$(1\ldots n)$ and $(\sigma_1\ldots\sigma_n)$ as in (2.18).  Then
$$
\sum_{j_1\ldots j_n}(-q)^{\ell(j)} t_{i_1j_1}\ldots t_{i_nj_n} =
(-q)^{\ell(i)}\Delta^q~. \eqno(7.8)
$$
\no When written to resemble the familiar formula for the usual determinant,
(7.8) becomes
$$
\sum_{j_1\ldots j_n} {\cal{E}}^q_{j_1\ldots j_n}
t_{i_1j_1}\ldots t_{i_nj_n} = {\cal{E}}^q_{i_1\ldots i_n} \Delta^q 
\eqno(7.9)
$$
\no where
$$
\eqalignno{{\cal{E}}^q_{j_1\ldots j_n} &= (-q)^{\ell(j_1\ldots j_n)} & (7.10a) \cr
\noalign{\hbox{and}}
{\cal{E}}^q_{j_1\ldots j_n} &= 0 \quad \hbox{unless all indices are
different.} & (7.10b) \cr}
$$

The ${\cal{E}}^q$ symbol thus behaves as an invariant tensor with weight
$\Delta^q$.  

The trilinear form
$$
I_3 = {\cal{E}}^q_{j_1j_2j_3} \varphi^{j_1}\varphi^{j_2}
\varphi^{j_3}
$$
\no is invariant under
$$
\varphi^{j^\prime} = \varphi^iT_i^j
$$
\no where the $\varphi^i$ commute with the $T_i^j$ since
$$
\eqalign{I_3^\prime &= {\cal{E}}^q_{j_1j_2j_3}
(\varphi^{j_1}\varphi^{j_2}\varphi^{j_3})^\prime \cr
&= {\cal{E}}^q_{j_1j_2j_3}(\varphi^{i_1}\varphi^{i_2}\varphi^{i_3}
T_{i_1}^{j_1}T_{i_2}^{j_2}T_{i_3}^{j_3}) \cr
&= ({\cal{E}}^q_{j_1j_2j_3}T_{i_1}^{j_1}T_{i_2}^{j_2}T_{i_3}^{j_3})
\varphi^{i_1}\varphi^{i_2}\varphi^{i_3} \cr
&= \Delta^q{\cal{E}}^q_{i_1i_2i_3}\varphi^{i_1}\varphi^{i_2}
\varphi^{i_3} \cr}
$$
\no or
$$
I_3^\prime = \Delta^qI_3 \eqno(7.11)
$$
\no by (7.9).

The invariance of the quantum determinant may also be used to form the inverse
of $T$ that is needed in (7.2).  By (7.9) with $\Delta^q=1$
$$
\sum_{j_1} t_{i_1j_1}\sum_{j_2\ldots j_n} {\cal{E}}^q_{j_1\ldots j_n}
t_{i_2j_2}\ldots t_{i_nj_n} = {\cal{E}}^q_{i_1\ldots i_n}
\eqno(7.12)
$$
\no i.e.
$$
\sum_{j_1}t_{i_1j_1}\sum_{j_2\ldots j_n} t_{i_2j_n}\ldots
t_{i_nj_n}(-q)^{\ell\left(\matrix{1~\ldots n\cr j_1\ldots j_n \cr}\right)}
= (-q)^{\ell\left(\matrix{1~\ldots n \cr i_1\ldots i_n \cr}\right)}~.
\eqno(7.13)
$$
\no Then
$$
\eqalign{
\sum_{j_1}t_{i_1j_1}\sum_{\scriptstyle j_2\ldots j_n \atop
\scriptstyle i_2\ldots i_n}t_{i_2j_2}\ldots t_{i_nj_n}
&(-q)^{\ell\left(\matrix{1~\ldots n\cr j_1\ldots j_n \cr} \right)}
(-q)^{-\ell\left(\matrix{12~\ldots n\cr si_2\ldots i_n \cr}\right)} \cr
&= \sum_{i_2\ldots i_n}(-q)^{\ell\left(\matrix{1~\ldots n\cr
i_1\ldots i_n \cr}\right)-\ell\left(\matrix{1~~\ldots n\cr
si_2\ldots i_n}\right)} \cr} \eqno(7.14)
$$
\no or
$$
\eqalign{
\sum_{j_1}t_{i_1j_1}\sum_{\scriptstyle j_1\ldots j_n \atop
\scriptstyle i_2\ldots i_n} t_{i_2j_2}\ldots t_{i_nj_n}
(-q)^{\ell\left(\matrix{si_2~\ldots i_n\cr
j_1j_2\ldots j_n \cr}\right)} &= \sum_{i_2\ldots i_n}
(-q)^{\ell\left(\matrix{si_2~\ldots i_n \cr i_1i_2\ldots i_n \cr}\right)} \cr
&= (n-1)!~\delta^s_{i_1} \cr}\eqno(7.15)
$$
\no since
$$
\ell\left(\matrix{1~\ldots n \cr j_1\ldots j_n \cr} \right) =
\ell\left(\matrix{1~\ldots n \cr i_1\ldots i_n \cr} \right) +
\ell\left(\matrix{i_1\ldots i_n \cr j_1\ldots j_n \cr} \right)~.
\eqno(7.16)
$$
\no Therefore the matrix inverse to $T$ is$^2$
$$
(T^{-1})_{j_1s} = {1\over (n-1)!}
\sum_{\scriptstyle i_2\ldots i_n \atop \scriptstyle j_2\ldots j_n}
t_{i_2j_2}\ldots t_{i_nj_n}
(-q)^{\ell\left(\matrix{si_2~\ldots i_n\cr j_1j_2\ldots j_n \cr}\right)}
\eqno(7.17)
$$
\vskip.5cm

\line{{\bf 8.~~Field Theory.} \hfil}
\vskip.3cm

In an earlier paper$^1$ the following field theoretic Lagrangian was
proposed
$$
{\cal{L}} = -{1\over 4}\sum_\alpha L(\alpha)\vec F_{\mu\nu}\vec F^{\mu\nu}
R(\alpha) + i~\tilde\psi C\epsilon\gamma^\mu\vec\nabla_\mu\psi +
{1\over 2}\bigl[(\tilde\varphi\buildrel\leftarrow\over
{\nabla}_\mu) \epsilon(\vec\nabla_\mu\varphi)
+ \tilde\varphi\epsilon\varphi\bigr]
\eqno(8.1)
$$
\no which is both Lorentz and $SU_q(2)$ invariant.  Here
$$
T^t\epsilon T = T\epsilon T^t = \epsilon \qquad
{\rm det}_qT = 1~.
$$
$$
\eqalign{L(\alpha)^\prime &= L(\alpha)T^{-1} \cr
R(\alpha)^\prime &= TR(\alpha) \cr
(L\epsilon)^\prime &= (L\epsilon)T^t \cr
\hfil \cr}
\eqalign{\varphi^\prime &= T\varphi \cr
\tilde\varphi^\prime &= \tilde\varphi T^t \cr
\psi^\prime &= T\psi \cr
\tilde\psi^\prime &= \tilde\psi T^t \cr} \quad
\eqalign{(\vec\nabla\varphi)^\prime &= T(\vec\nabla\varphi) \cr
(\tilde\varphi \buildrel \leftarrow\over\nabla)^\prime &=
(\varphi \buildrel\leftarrow\over\nabla)T^t \cr
(\vec\nabla\psi)^\prime &= T(\vec\nabla\psi) \cr
(\tilde\psi \buildrel\leftarrow\over\nabla)^\prime &=
(\tilde\psi\buildrel\leftarrow\over\nabla)T^t \cr}
\eqno(8.2)
$$
\no Kinetic terms in $L(\alpha)$ and $R(\alpha)$, as well
as other possible terms in $\buildrel\leftarrow\over A$
have not been expressed in (8.1).

The invariance of the Lagrangian requires distinct left
and right fields since $F$ does not commute with $T$.  In
the limit $q=1$ of (4.2) the $L$ and $R$ fields may be
summed out as follows:
$$
\sum_\alpha L_i(\alpha)R_j(\alpha) = \delta_{ij}
$$
\no where the sum is over the complete set of left and
right fields.  Then
$$
\lim_{q\to 1} \sum_{\alpha} L_i(\alpha)(FF)_{ij}
R_j(\alpha) = {\rm Tr}~FF~. \eqno(8.3)
$$

To pass from (8.1) to Lagrangians with symmetries of 
higher rank, one may replace the bilinear invariants
$\tilde\varphi\epsilon\varphi$ and $\tilde\psi C\epsilon
\nabla\!\!\!/ \psi$ by $\tilde\varphi R\varphi$ and
$\tilde\psi CR\nabla \!\!\!/\psi$ where the invariance of
the $R$-forms is described in Eqs. (7.4)-(7.6).

Now assume that the generic field lies in the algebra:
$$
\eqalignno{\psi_{k\ell}(x) &= \sum\varphi_{k\ell}^{\alpha\beta}
t_{\alpha\beta} & (8.4) \cr
\tilde\psi_{mp}(x) &= \sum t_{\alpha\beta}
\tilde\varphi_{mp}^{\alpha\beta}~, & (8.5) \cr}
$$
\no where the $\varphi$ do not lie in the algebra.
In a general gauge transformation
$$
\eqalignno{\tilde\psi^\prime_{mp}(x) &= \sum\tilde\psi_{ij}(x)
T(2)_{jp} T(1)_{im} & (8.6) \cr
\psi^\prime_{k\ell}(x) &= \sum(T(2)^{-1})_{\ell s}
(T(1)^{-1})_{kt}\psi_{ts}(x) ~. & (8.7) \cr}
$$
\no Then
$$
\sum\tilde\psi_{ij}(x)R_{ijts}\psi_{ts}(x) =
\sum\tilde\psi^\prime_{mp}(x) R_{mpk\ell}\psi^\prime_{k\ell}(x)~.
\eqno(8.8)
$$

Distinguish between the $H(i,i)$ and the $E(i,j)$ fields by associating
the $H(i,i)$ with a Lorentz spinor field and $E(i,j)$ with a Lorentz
vector field in a particular gauge. In this special gauge, one may
write the spinor field as
$$
\psi = \sum^3_{i=1}\varphi^\alpha H(\alpha)~, \quad
\alpha = 1,2,3 \eqno(8.9)
$$
\no and a mass term as
$$
M \tilde\psi CR\psi = M\sum^3_{i=1} H(\alpha)^2 \eqno(8.10)
$$
\no if we orthonormalize as follows:
$$
\tilde\varphi^\alpha CR\varphi^\beta = \delta^{\alpha\beta} \eqno(8.11)
$$
\no Here $\tilde\psi$ is a transposed spinor and $C$ is the charge
conjugation matrix.

The eigenvalue of this term in the state $|n_1n_2n_3\rangle$ is by (4.6)
$$
M\sum^3_{i=1} q^{-2n_i} \eqno(8.12)
$$
\no and the masses associated with the three-spinor fields are
$$
m_i = q^{-2n_i} M~. \eqno(8.13)
$$
\no If the field particles are all in their ground states, then by (6.6)
their masses are related by
$$
m_1m_2m_3 = M^3q_1^6~. \eqno(8.14)
$$
\no Each field particle may also exist in excited states according to
(8.13).

These three spinor fields interact through terms like
$$
\tilde\psi CR \nabla \!\!\!\!/\psi \eqno(8.15)
$$
\no where
$$
\nabla_\mu = \partial_\mu + A_\mu \qquad
\nabla \!\!\!\!/ = \gamma^\mu\nabla_\mu~. \eqno(8.16)
$$

The
$q$-invariance of the interaction term requires
$$
\tilde\psi^\prime CR\nabla\!\!\!\!/^\prime\psi^\prime =
\tilde\psi CR\nabla\!\!\!\!/\psi~. \eqno(8.17)
$$

Let us abbreviate (8.6) and (8.7)
$$
\eqalignno{\tilde\psi^\prime &= \tilde\psi
\buildrel \leftarrow\over {\cal{T}} & (8.18) \cr
\psi^\prime &= \vec{\cal{T}}\psi & (8.19) \cr}
$$
\no We shall also assume
$$
\eqalignno{(\nabla\!\!\!\!/\psi)^\prime &= \vec{\cal{T}}
(\nabla\!\!\!\!/\psi) & (8.20) \cr
\noalign{\hbox{and therefore}}
\nabla\!\!\!\!/^\prime \vec{\cal{T}} &= \vec{\cal{T}} \nabla\!\!\!\!/ & (8.21) \cr}
$$
\no then
$$
\tilde\psi^\prime CR(\nabla\!\!\!\!/\psi)^\prime = \tilde\psi CR
(\nabla\!\!\!\!/\psi)~. \eqno(8.22)
$$
\no A natural choice for $A$ is
$$
A\!\!\!/ = \sum^3_{i=1} (A\!\!\!/(i)E(i) + \bar A\!\!\!/(i)
\bar E(i))~. \eqno(8.23)
$$
\no Then by (8.9) and (8.23) the invariant interaction (8.17) induces
the following probability amplitude for the transition
$\vec n\to \vec n^\prime$:
$$
\eqalignno{&\langle n_1^\prime n_1^\prime n_3^\prime|\tilde\psi CR
A\!\!\!/\psi|n_1n_2n_3\rangle \cr
&= \langle n_1^\prime n_2^\prime n_3^\prime|\sum^3_{i=1}
(\tilde\psi(i)H(i))CR(\sum^3_{j=1}A\!\!\!/(j)E(j)|(\sum^3_{k=1}\psi(k)H(k))
|n_1n_2n_3\rangle \cr
& ~~~~+ \hbox{contribution of}~\bar A\!\!\!/i)\bar E(i) & (8.24) \cr
&= \sum_{ijk}[\tilde\psi(i)CRA\!\!\!/(j)\psi(k)]
\langle n_1^\prime n_1^\prime n_3^\prime|H(i)E(j)H(k)|n_1n_2n_3\rangle \cr
& ~~~~+ \hbox{contribution of}~\bar A\!\!\!/(i)\bar E(i) \cr
&= \sum_{ijk}[\tilde\psi(i)CRA\!\!\!/(j)\psi(k)]q_1^{n_i+n_k}\alpha_i\alpha_k
\langle n_1^\prime n_2^\prime n_3^\prime|E(j)|n_1n_2n_3\rangle \cr
& ~~~~+ \hbox{contribution of}~\bar A\!\!\!/(i)\bar E(i) & (8.25) \cr}
$$\no The sum on $(ijk)$ is restricted by (4.10) to the following
combinations
$$
\eqalign{&j=1 \qquad (i,k) = 1,2,3 \cr
&j=2 \qquad (i,k) = 1,2 \cr
&j=3 \qquad (i,k) = 2,3 \cr} \eqno(8.26)
$$
\no It is also limited by the following selection rules
$$
\eqalign{\langle\vec n^\prime|\bar E(1)|\vec n\rangle &= 0 \cr
\langle\vec n^\prime|\bar E(2)\vec n\rangle &= 0 \cr
\langle\vec n^\prime|\bar E(3)|\vec n\rangle &= 0 \cr}\quad
\eqalign{\hbox{unless} \hfil \cr \hfil \cr \hfil \cr} \quad
\eqalign{n_1^\prime &= n_1+1 \cr n_1^\prime &= n_1+1 \cr
n_1^\prime &= n_1 \cr} \quad
\eqalign{n_3^\prime &= n_3+1 \cr n_2^\prime &= n_2+1 \cr
n_2^\prime &= n_2+1 \cr} \quad
\eqalign{n_2^\prime &= n_2+2 \cr n_3^\prime &= n_3 \cr n_3^\prime &= n_3+1 \cr}
\eqno(8.27)
$$
\no When $q=1$ one has simply
$$
\tilde\psi CA\!\!\!/ \psi~. \eqno(8.28)
$$
\no The new factors when $q\not= 1$ include 
$$
q_1^{n(i)+n(j)}\alpha_i\alpha_j = (m_im_j)^{1/2}/M \eqno(8.29)
$$
\no as well as a similar factor from $\langle\vec n^{~\prime}|E|\vec n\rangle$,
together
implying that the associated probabilities for absorption and emission
of a pair of heavy spinor particles by the vector particles proportional to
the (product)$^2$ of the spinor masses.

We have described the elements of a formalism resembling the current
non-Abelian gauge theories.  The Lagrangian of this model is built out of
the three $H$-fields and the three $E$-fields.  If the $H$-fields are spinor
fields, they are quarklike and if the three $E$-fields are gauge vectors,
then they resemble gluons.  If the interaction between the ``quarks" and
``gluons" is given by (8.15), then according to the selection rules (8.27)
the ``quarks" cannot be singly excited.

When the Lie group is deformed, the group $(g)$ and the group
algebra $(u)$ undergo quite different deformations; and as a result the full
deformed theory contains two sectors, one coming from $g$ and the other 
from $u$.  In this paper we have discussed the $g$-sector.  The particles
lying in this sector become invisible in the $q=1$ limit and are nearly
invisible if $q$ is close to this limit.  The $u$-sector, which we have not
discussed here, results from the deformation of the algebra $(u)$. 
In the $SU_q(2)$ case the deformed generators $(J^q_{(\pm)},J^q_{(3)})$
satisfy
$$
\eqalign{(J_3^q,J_\pm^q) &= \pm J_\pm^q \cr
(J_+^q,J_-^q) &= {1\over 2}[2J_3^q]_q \cr
[2J_3^q]_q &= {q^{2J_3^q}-q^{-2J_3^q}\over q-q^{-1}}~. \cr}
\eqno(8.30)
$$
\no As $q\to 1$ these relations approach the corresponding relations for the
Lie algebra so that the usual Yang-Mills theory is recovered.  The
deviation of the $u$-sector from the standard theory 
is accessible by perturbation theory if $q$ is near unity.

More interesting, however, if $q$ is near unity, is the $g$ sector which
contains the hypothetical, exotic particles discussed in this paper and
which are nearly invisible if $q$ is near unity.  We have
not discussed the $u$-sector in this paper.

\vskip.5cm

\line{{\bf References.} \hfil}
\vskip.3cm

\item{1.} R. Finkelstein, hep-th/0003189.
\item{2.} N. Ya. Reshetikhin, L. A. Takhtadzhan, and L. D. Fadeev,
Leningrad Math. J{\bf 1} (1990) No. 1.

\end
\bye